**Thermal Effects on the Magnetic Field Dependence of Spin Transfer Induced Magnetization Reversal**

D. Lacour, J. A. Katine*, N. Smith, M. J. Carey, J. R. Childress

*Hitachi Global Storage Technologies San Jose Research Center*

*650 Harry Rd., San Jose, CA 95120*

## Abstract

We have developed a self-aligned, high-yield process to fabricate CPP (current perpendicular to the plane) magnetic sensors of sub 100 nm dimensions. A pinned synthetic antiferromagnet (SAF) is used as the reference layer which minimizes dipole coupling to the free layer and field induced rotation of the reference layer. We find that the critical currents for spin transfer induced magnetization reversal of the free layer vary dramatically with relatively small changes the in-plane magnetic field, in contrast to theoretical predictions based on stability analysis of the Gilbert equations of magnetization dynamics including Slonczewski-type spin-torque terms. The discrepancy is believed due to thermal fluctuations over the time scale of the measurements. Once thermal fluctuations are taken into account, we find good quantitative agreement between our experimental results and numerical simulations.

* Electronic address: jordan.katine@hgst.com

Since its first experimental observation five years ago[1,2], the spin transfer effect has intrigued researchers with its potential for application to magnetic random access memory (MRAM)[3] and high frequency, current-driven oscillators.[4,5] The spin transfer phenomenon can be most readily exploited in current-perpendicular-to plane (CPP) devices with lateral dimensions on the order of 100 nm, a regime where processing variations, the granular nature of polycrystalline magnetic thin films, and dipole-coupling to the pinned layer(s) will inevitably create substantial variations in the effective anisotropy and in-plane magnetic fields experienced by the free layer of such devices. To assess the feasibility and/or nature of spin transfer in applications, a self-aligned, high yield process has been developed to produce highly uniform, sub-100 nm, CPP GMR spin-valve devices.

Device fabrication begins with sputter deposition of (thicknesses in nm): substrate/bottom lead/antiferromagnet/pinned layer/Cu (4.0)/free layer/cap, where the bottom lead is a Ta (0.5)/Cu (20)/Ta (3)/Cu (20)/Ta (2.5) film, the antiferromagnet is PtMn (17.5) which exchange couples to the synthetic antiferromagnet (SAF) pinned layer consisting of $Co_{80}Fe_{20}$ (1.5)/Ru (0.8)/ $Co_{80}Fe_{20}$ (1.9). The free layer consists of $Co_{80}Fe_{20}$ (1.0)/$Ni_{86}Fe_{14}$ (2.4). The cap, Cu (20)/Ru (5)/Ta (2.5), protects the free layer from oxidation during the annealing process, and allows good electrical contact to be made to the top of the device. Measurements show that the exchange coupling between the PtMn and the SAF keeps the second, 1.9 nm CoFe *reference* layer well pinned to applied in-plane fields up to 2 kOe. The SAF also substantially reduces the dipole field on the free layer from the reference layer, allowing easier quantitative examination of spin-transfer effects in relatively small in-plane fields.

Following film deposition, electron beam lithography is used to pattern HSQ, a high resolution negative e-beam resist spun approximately 100 nm thick.[6] Electron beam exposure essentially converts HSQ into $SiO_2$, whose excellent resistance to ion milling allows the resist to directly serve as a high-fidelity mask during the etching of our devices. Figure 1 shows top down SEM images of the 3 fabricated devices (after ion milling) with hexagonal shapes and short (hard) axis lengths of 50, 75, and 100 nm. The long (easy) axis of the hexagon is parallel to the pinning direction. After milling, 100 nm of aluminum oxide is ion beam deposited onto the wafer, which is then briefly chemically mechanically polished (CMP). Due to the mechanical instability of the resist mask, it is easily removed during the CMP process, creating a via in the aluminum oxide that allows self-aligned contact to the top of pillar.

The inset to Fig. 2b shows a magnetic hysteresis loop from a 50 nm hexagon. It was measured at room temperature, as were all other measurements reported herein. An applied field along the $\hat{e} \equiv$ easy axis is defined as positive when *antiparallel* to the fixed reference layer magnetization, and favors the high resistance (antiparallel) alignment. The openness of the hysteresis loop defines the *coercive field*, $H_c$, and the curve is offset by the *dipole field* $H_d \cong +60$ Oe from the SAF on the free-layer. From the lithographic area and the measured hysteresis loops, we plot the RA product vs. ΔR/R for 36 hexagonal devices in Fig. 2a. The data are very well clustered, indicative of the small variance of our sub-100 nm fabrication process. The ΔR/R value is lower than that published in earlier unpinned samples primarily because of the large resistance of the PtMn layer.[3] The ΔRA product of 0.35 mΩ.µm$^2$ is ~20% lower than observed in identical devices with a simple $Co_{90}Fe_{10}$ pinned layer, possibly due to a reduction in the net polarization, *P*, of the current emerging from a SAF pinned layer.[7]

Despite the relatively small spread in RA, ΔR/R, and $H_d$, Fig. 2b shows considerable spread in coercivity $H_c$. The main contributors to the measured coercive field $H_c$ are the combined crystalline plus shape anisotropy, the latter being additionally sensitive to the lithographic variations in the difference in length for the long and short axes of the hexagons. As discussed below, the temperature, device volume, and dwell time for the measurement will also have a substantial influence.

Plotted in Fig. 3a, is the differential resistance, dV/dI vs. the *dc* current bias, $I_b$. A lock-in amplifier and 10 μA excitation current were used to measure the differential resistance. The *dc* current was stepped in 100 μA increments, with a *one second* dwell time. By definition, positive bias current $I_b > 0$ means electrons flow from the free layer towards the reference/pinned layers. The curve of zero bias dV/dI vs. applied field $H_a$ for this $H_c$ =80 Oe, $H_d$=60 Oe device is pictured in the inset of Fig. 2b. At $H_a$=-90 Oe, corresponding to the *net* in-plane field $H_{fre} \equiv H_a + H_d$ =-30 Oe on the free layer, the device starts in the parallel (low resistance) orientation at zero current, switches to anti-parallel alignment at a critical current $I_c^+$ =1.5 mA., and remains in that state until the bias current is swept below a second critical current $I_c^-$=-0.3 mA. The polarity of this current-induced switching is consistent with the spin transfer effect. When the dV/dI vs. $I_b$ loops are repeated with increasing values of $H_a$, $I_c^+$ steadily decreases, while $I_c^-$ shifts very little, leading to a substantial narrowing of the hysteresis-width in current. When $H_a$ =30 Oe ($H_{fre}$=90 Oe), the bias current hysteresis is virtually gone, and the current-induced switching appears completely reversible at $H_a$ =60 Oe ($H_{fre}$=120 Oe).

Phase diagrams for this and two other (lower $H_c$) 50 nm hexagons are plotted in Fig. 3, showing $I_c^+$ and $I_c^-$ as functions of the total field $H_{fre} = H_a + H_d$ on the free layer. In all devices, $I_c^+$ decreases much more quickly with increasing field than does $I_c^-$, eventually leading to a crossover after which the switching appears reversible. A similar crossover between reversible and hysteretic switching in unpinned Co/Cu/Co devices was observed at low temperature (30 K) by Grollier, et al.[8] However, their observations occurred in a "high-field" regime ($H_a$ =0.5 kOe) such that $H_a$ >> $H_k$, where $H_k$ is the net (shape plus crystalline) *intrinsic* uniaxial anisotropy (in-plane) of the free-layer.

Provided that $|H_{\mathrm{fre}}| < H_k$, a stability analysis[8,9] of the (linearized) Landau-Lifshitz-Gilbert (LLG) equations of motion for the free layer magnetization, including the Slonczewski form[10] of the spin-transfer torques, yields the following predictions:

$$I_{c0}^{\pm} = (3\times 10^{11}\ \text{mA/Oe/emu})\,\alpha\, M_s V (H_k + H_{\mathrm{fre}} + H_{\perp}/2)/(\nu^{\pm} P) \qquad (1)$$

for the critical switching currents at *low* temperature. Here, $M_s$, $V$, $H_{\perp}$, and $\alpha$ are the saturation magnetization, volume, *out-of-plane* anisotropy, and Gilbert damping constant for the free layer, and $\nu^{\pm}$ is a polarization-dependent factor.[10,11] Since $H_{\perp} \sim 4\pi M_s \sim 10$ kOe >> $H_{\mathrm{fre}}$ and/or $H_c$, Eq. (1) cannot explain the observed large variations in the *measured* $I_c^{\pm}$ with changes in $H_{\mathrm{fre}}$ of merely several tens of Oe.

Equation 1 excludes explicit consideration of *thermally induced* fluctuations, which have earlier been shown to reduce the critical currents.[12,13] In the present context, thermally activated magnetization reversal of the free-layer will firstly reduce the measured coercivity $H_c$ relative to the intrinsic $H_k$. This effect is well known,[14] and treating the free-layer as a single domain uniaxial magnet, can be expressed as

$$H_c(H_k) \cong H_k \{1 - [k_B T / E_k \ln(f_0 \tau_m / \ln 2)]^{1/2}\} \qquad (2)$$

where intrinsic energy barrier $E_k = \frac{1}{2} M_s V H_k$, $f_0 \sim 10^9$ Hz is an "attempt frequency",[15] and $\tau_m$ >> $1/f_0$ is the measurement time. For the 50 nm ($M_s V \cong 1.1\times 10^{-14}$ emu), $H_c \cong 80$ Oe, $H_d \cong 60$ Oe device described in Fig. 2b-inset and Fig. 3a, one can use Eq. 2, assuming $T$ =310K and $\tau_m$ =dwell-time = 1 sec, to estimate an intrinsic $H_k \approx 300$ Oe, (primarily shape anisotropy of the hexagonal pillar). That $H_k >> H_c$ indicates the important role of thermal activation in the present experiment. Additionally, thermal fluctuations of the free-layer magnetization in the presence of the *non-conservative* spin-transfer torques can lead to additional absorption of energy.[9] The latter was shown to result in a bias current dependent effective anisotropy $H_k^{\mathrm{eff}}(I_b, H_{\mathrm{fre}}; H_k)$, with corresponding influence on the energy barrier $E_k(H_k^{\mathrm{eff}})$ for easy-axis reveral.[16] This interpretation is consistent with other recent measurements.[17]

These effects were analyzed here by numerical solution of the thermal-field plus spin-torque-augmented LLG equations for a single-domain free layer:

$$
\begin{aligned}
&(1+\alpha^2)\frac{d\hat{m}_{\mathrm{fre}}}{dt} = \gamma(\vec{H}_{\mathrm{eff}} \times \hat{m}_{\mathrm{fre}}) + \alpha\gamma(\hat{m}_{\mathrm{fre}} \times \vec{H}_{\mathrm{eff}} \times \hat{m}_{\mathrm{fre}}) \\
&\vec{H}_{\mathrm{eff}} = \vec{H}_{\mathrm{fre}} + \vec{H}_{\mathrm{th}}(t) + H_k'(\hat{m}_{\mathrm{fre}} \cdot \hat{e})\hat{e} + H_\perp(\hat{m}_{\mathrm{fre}} \cdot \hat{\perp})\hat{\perp} + H_{\mathrm{st}}(\hat{m}_{\mathrm{ref}} \times \hat{m}_{\mathrm{fre}})
\end{aligned} \tag{3}
$$

where spin-torque field $H_{\mathrm{st}} = (3.3\times10^{12}\,\mathrm{Oe\text{-}emu/mA})\nu(\hat{m}_{\mathrm{fre}} \cdot \hat{m}_{\mathrm{ref}})PI_b/(M_sV)$.[9,15] At each time step of $\Delta t$ = 2 psec, the Cartesian components of random thermal field $\vec{H}_{\mathrm{th}}$ were chosen as zero-mean gaussian random numbers with $\sigma_H = \sqrt{4k_BT/\gamma M_sV\Delta t}$.[9,14] The free layer used in these CPP devices is similar to that used in Hitachi's commercial GMR spin valves, whose well characterized properties include $\alpha$ =0.02, $H_\perp$ =8.5 kOe, and $M_st_{\mathrm{fre}}$ =0.32 memu/cm$^2$, and $T$=310 K. The polarization $P$ was assumed to be 0.3. For practical computational constraints, the bias current $I_b(t)$ was chosen to be a 10 kHz triangle wave rather than the ~0.01 Hz sweep rate for the data of Fig.3a, and the results for $\hat{m}_{\mathrm{fre}}(t)$, were computed as sequential averages $\langle m_{\mathrm{fre}}(t)\rangle$ over contiguous time intervals of 0.4 μsec. To compensate, the simulation assumes a value $H_k'$ =170 Oe, chosen via Eq. (2) with $\tau_m$ =0.4 μsec to predict the same thermally activated value of $H_c$=80 Oe as was measured for the 50 nm device of Fig. 3a. The simulated *dV/dI* loops (with *dV/dI* taken $\propto -\langle m_{\mathrm{fre}}(t)\rangle \cdot \hat{m}_{\mathrm{ref}}$) are shown in Fig.3c. Both qualitatively and quantitatively, the agreement with the Fig. 3a data is remarkably good, perhaps somewhat fortuitously so given uncertainty in both $P$ and $\nu$.[11]

To further illustrate how small changes in $H_c$ can have a large effect on the critical switching currents, the critical current densities at $H_{fre}$ =0 as a function of $H_c$ are plotted in Fig. 4. As the coercivity/anisotropy of these samples increases, the critical current densities required for switching also substantially increase, and climb above $10^7$ A/cm$^2$. Also shown in Fig. 4 is a model prediction,[16]

$$I_c^{\pm}(H_c) = I_{c0}^{\pm}\,[1 - H_k(H_c \to 0^+)/H_k(H_c)] \qquad (4)$$

where $H_k(H_c)$ is the (numerical) *inverse* function of Eq. (2) with $\tau_m = 1\,\mathrm{sec}$, and $I_{c0}^{\pm}$ is from Eq. (1) with device parameter values the same as previously quoted, with the exception that $P$ =0.2 here. The agreement for 50 and 75 nm devices is reasonably good for both polarities $J_c^{\pm}$.

In summary, our study of pinned-SAF pillar devices demonstrates that the threshold values for current-induced switching vary dramatically with the effective in-plane magnetic field experienced by the free layer. Analytical models and numerical simulations seem to confirm that this strong field dependence stems from the role thermal excitations play in assisting the spin transfer induced magnetization reversal. Comparison of data and model also appears to confirm the substantial polarity-asymmetry of the critical switching currents polarization $I_c^{\pm}$ predicted theoretically.[10,11]

We acknowledge C. T. Black, B. A. Gurney, and J. Li for their assistance in these experiments.

## Figure Captions

Figure 1. SEM images of 50, 75, and 100 nm hex devices. Respective areas are .0035, .0083, and .0138 $\mu m^2$.

Figure 2. (a) Scatter in $\Delta R/R$ vs. RA product. The results for all device sizes are essentially identical after a constant lead resistance of 1.8 $\Omega$ is subtracted from the raw data. (b) Scatter in the dipole and coercive fields $H_d$ and $H_c$. The inset showing a zero-bias hysteresis loop from a 50 nm devices shows how $H_d$ and $H_c$ are defined.

Figure 3. (a) $dV$/dI vs $I_b$. The critical currents for reversing the free layer magnetization vary with the applied in-plane field. (b) Phase diagram illustrating how the critical switching currents vary with $H_{fre}$ for several 50 nm devices. (c) Results from numerical simulations for different values of $H_a$.

Figure 4. The critical current densities for current-induced reversal increase with increasing coercive field. Solid lines are calculated from Eq. 4 assuming P = 0.2.

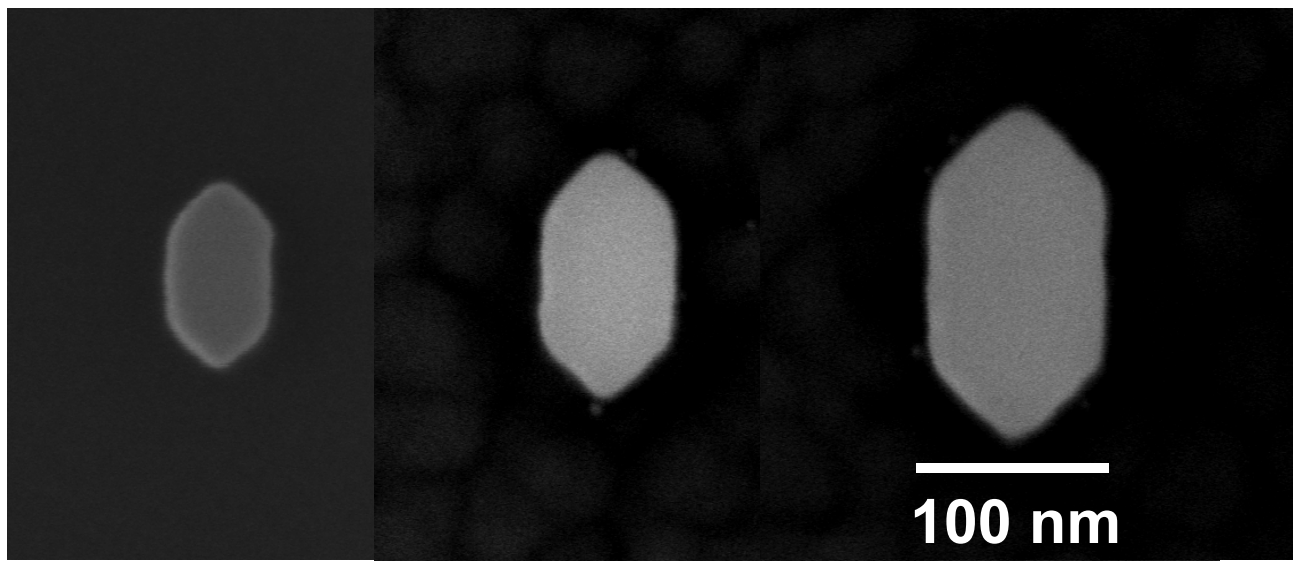

Fig 1. Lacour et al.

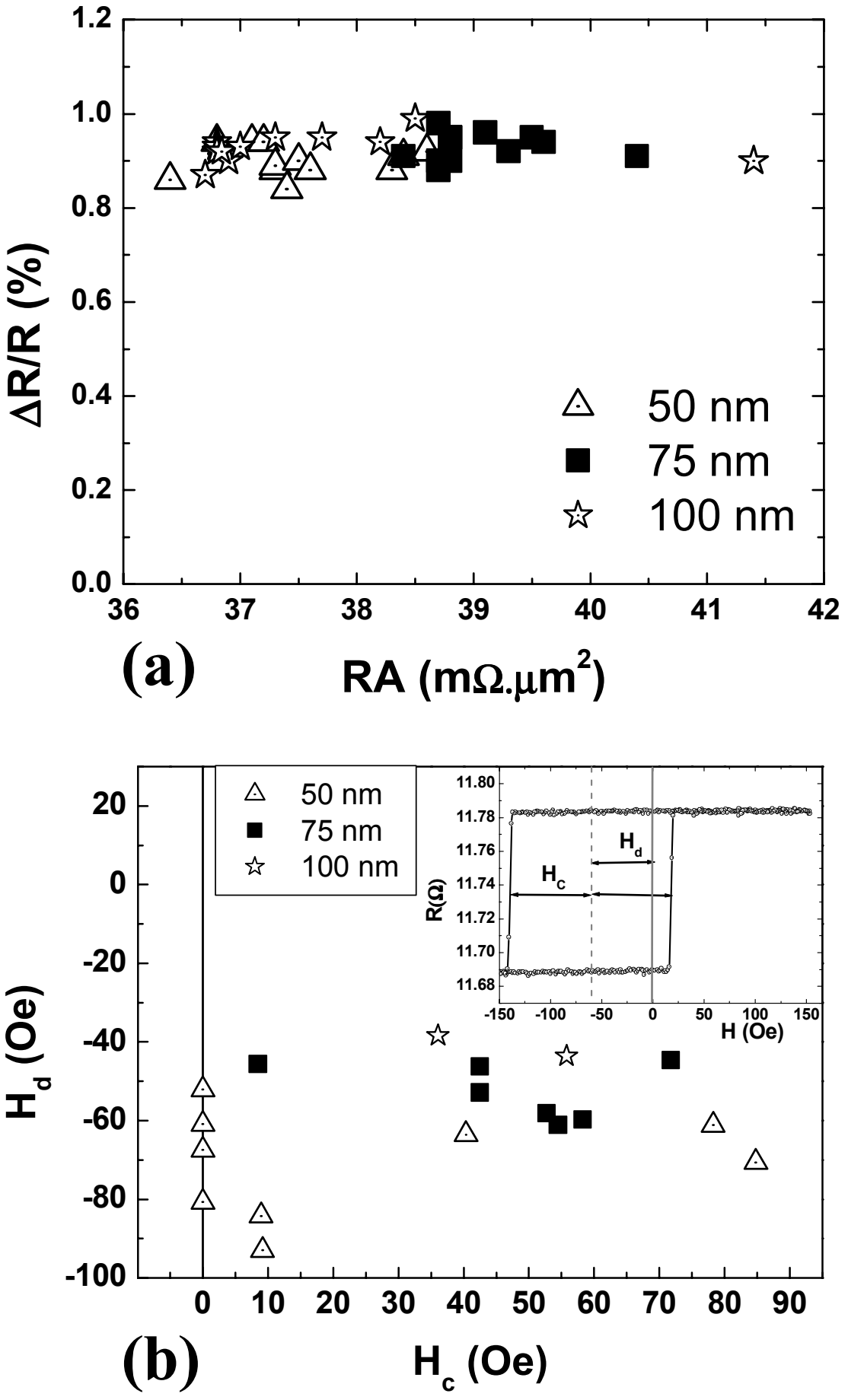

Fig 2. Lacour et al.

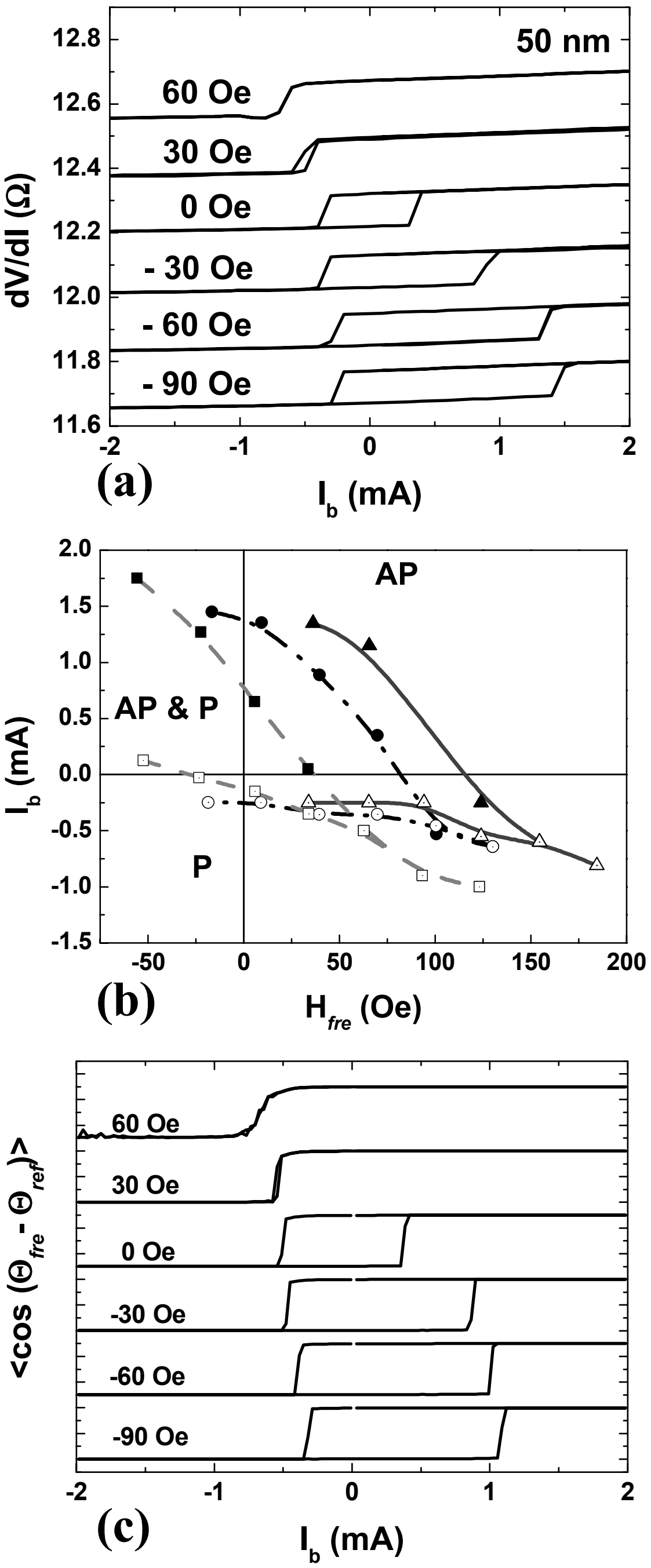

Fig 3. Lacour et al.

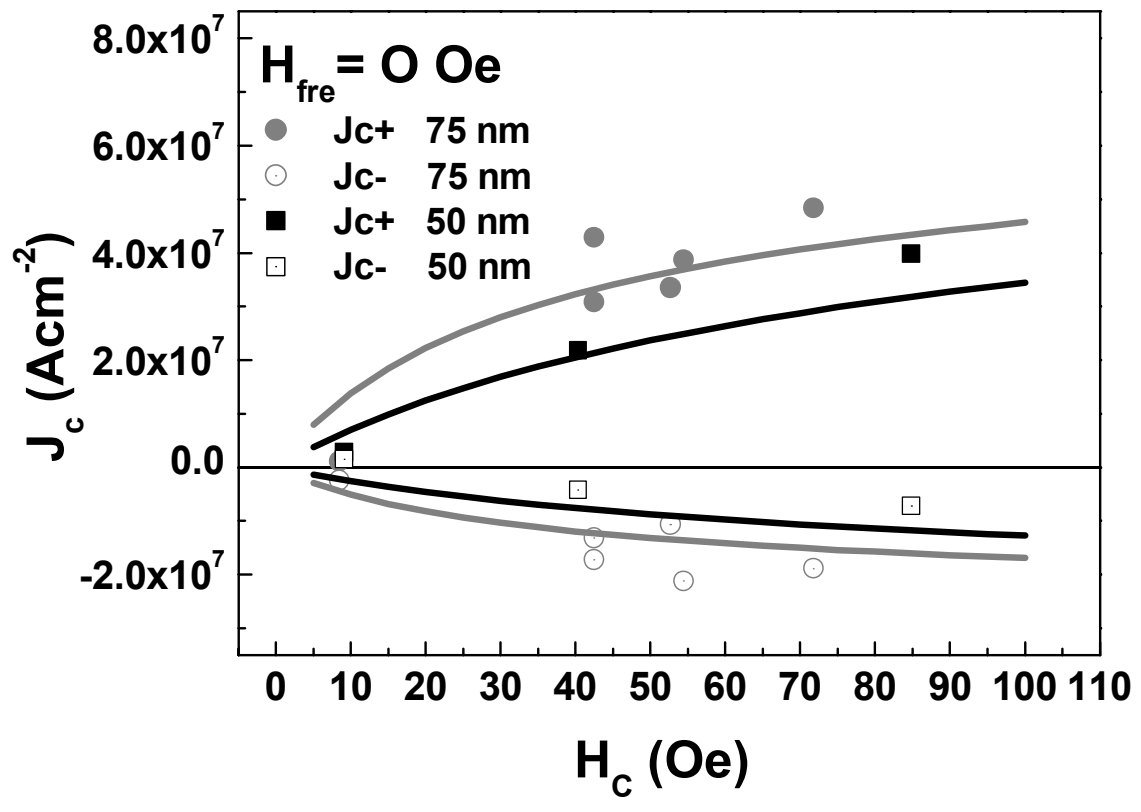

Fig 4. Lacour et al.